\documentclass{aa}
\usepackage{graphics}
\begin{document}

\title{Large-scale magnetized outflows from the Virgo Cluster spiral NGC\,4569}

\subtitle{A galactic wind in a ram pressure wind}

\author {
K. T. Chy\.zy\inst{1}
\and M. Soida\inst{1}
 \and D.J. Bomans\inst{2}
 \and B. Vollmer\inst{3}
 \and Ch. Balkowski\inst{4}
 \and R. Beck\inst{5}
 \and M. Urbanik\inst{1}}
\institute{Astronomical Observatory, Jagiellonian
University, ul. Orla 171, 30-244 Krak\'ow, Poland
\and Astronomisches Institut, Ruhr-Universit\"at-Bochum,
44780 Bochum, Germany
\and CDS, Observatoire astronomique de Strasbourg, UMR                    
7550, 11 rue de l'Universit\'e, 67000 Strasbourg,                           
France
\and Observatoire de Paris, GEPI, CNRS UMR 8111, and Universit\'e Paris 7, 5 Place
Jules Janssen, 92195 Meudon Cedex,  France
\and Max-Planck-Institut f\"ur Radioastronomie, Auf dem
H\"ugel 69, 53121 Bonn, Germany}

\offprints{K.T. Chy\.zy}
\mail{chris@oa.uj.edu.pl}
\date{Received date/ Accepted date}

\titlerunning{Large-scale magnetized outflows in the H{\sc i} deficient 
Virgo cluster spiral galaxy NGC\,4569}
\authorrunning{K.T. Chy\.zy et al.}

\abstract{
Using the Effelsberg radio telescope at 4.85~GHz and 8.35~GHz we 
discovered large symmetric lobes of polarized radio emission around the 
strongly H{\sc i} deficient Virgo cluster spiral galaxy NGC\,4569.
These lobes extend up to 24~kpc from the galactic disk. Our 
observations were complemented by 1.4~GHz continuum emission from 
existing H{\sc i} observations. This is the first time that such 
huge radio continuum lobes are observed in a cluster spiral galaxy.
The eastern lobe seems detached and has a flat spectrum typical for 
in-situ cosmic ray electron acceleration. The western lobe is 
diffuse and possesses vertical magnetic fields over its whole volume. 
The lobes are not powered by an AGN, but probably by a nuclear 
starburst producing $\geq 10^5$ supernovae which 
occurred $\sim 30$~Myr ago. Since the radio 
lobes are symmetric, they resist ram pressure due to the galaxy's 
rapid motion within the intracluster medium.  
}

\maketitle

\section{Introduction \label{sec:intro}}

NGC\,4569 is a bright spiral galaxy (Sb) whose projected angular distance to 
the  Virgo Cluster center (M87) is only $1.7^{\circ} = 0.5$~Mpc\footnote{We use 
a distance to the Virgo cluster of $D=17$~Mpc.}. Because of it brightness and 
large diameter ($D_{25}=9.5' = 47$~kpc) Stauffer et al. (\cite{stauffer}) have 
questioned its cluster membership. NGC\,4569 has about one tenth the H{\sc i} 
content of a field galaxy of the same morphological type and the same size 
(Giovanelli \& Haynes \cite{giovanelli}). It shows a strongly truncated
H{\sc i} disk (Cayatte et al. \cite{cayat1}), most probably a signature of 
strong stripping by the intracluster medium which pervades the Virgo cluster 
(Cayatte et al. \cite{cayat2}). Tsch\"oke et al. (\cite{tschoke}) did not 
find any soft X-ray emission (0.1--0.4 keV) from the northern half of the disk, 
whereas it is pronounced in the southern disk (cf. Fig~\ref{tp6}). These 
findings are consistent with a ram pressure scenario where the galaxy is 
moving to the north-east through the intracluster medium. The H$\alpha$ 
emission distribution is, as the H{\sc i}, sharply truncated at 30\% of the 
optical radius (Koopmann et al. \cite{koopmann}). In addition, an anomalous 
H{\sc i} and H$\alpha$ arm is detected to the west of the galactic disk.
Simulated H{\sc i} gas distributions and velocity fields of a more edge-on 
ram pressure stripping event (the inclination angle between the disk and 
the orbital plane is $35^{\circ}$), where the maximum ram pressure occurred 
$\sim 300$~Myr ago, are consistent with the H{\sc i} observations (Vollmer 
et al. \cite{vollm2}). 

Tsch\"oke et al. (\cite{tschoke}) discovered a diffuse extraplanar region 
of X-ray and H$\alpha$ emission to the west of the galactic disk. This 
was the first evidence of a direct connection between the hot X-ray gas 
and the H$\alpha$ emission at scales of 10~kpc. This ionized gas is flowing 
out from the disk at a velocity of 120 km s$^{-1}$ (Bomans et al. 
\cite{bomans05}). The most probable source of this outflow is a central 
starburst (Barth \& Shields \cite{barth00}, Tsch\"oke et al. \cite{tschoke}). 
Present-day AGN activity is ruled out due to the lack of a compact point 
source in the ASCA hard band (Tsch\"oke et al. \cite{tschoke}), a missing 
radio continuum point source (Neff \& Hutchings \cite{neff}, Hummel et al. 
\cite{hummel}), and spectral synthesis analysis (Barth \& Shields 
\cite{barth00}). Gabel \& Bruhweiler (\cite{gabel}) dated the nuclear 
starburst in the inner $\sim 30$~pc to $5-6$~Myr based on optical and UV
HST spectra. In addition, Keel (\cite{keel}) found a more extended ($\sim 300$~pc) 
population of A-type supergiants whose age is greater than $\sim 15$~Myr. Thus 
the nucleus of NGC\,4569 contains at least two distinct young stellar 
populations: a very young UV core ($5-6$~Myr) and a spatially more extended 
region dominated by A supergiants.

In this work we use radio polarimetry to study the gas outflows from NGC\,4569.
Total power emission and, first of all, polarized radio continuum emission 
represent a very sensitive tracer of flows of diffuse magnetized gas, often 
unnoticed in the H{\sc i} and H$\alpha$ lines or in X-rays (Beck et al. 
\cite{beck99}, Soida et al. \cite{soi4414}). While the polarization allows 
to study the magnetic field in the plane of the sky, the Faraday rotation 
measures the sign and strength of the magnetic fields along the line-of-sight. 
The knowledge of the spectral index variations across radio continuum 
structures allows to trace the history of relativistic electron population 
transported with the gas and magnetic fields. In the absence of an in-situ 
acceleration mechanism the relativistic electrons lose their energy via 
synchrotron emission. Since electrons with higher energies lose their energy 
more rapidly, the radio spectrum steepens with time (electron aging). 
To obtain maps of Faraday rotation measures and spectral index we performed 
observations at two frequencies:  4.85~GHz and 8.35~GHz. Additionally we used 
the H{\sc i} continuum at 1.4~GHz obtained by Vollmer et al. (\cite{vollm2}). 
The observations and the data reduction are presented in 
Sect.~\ref{sec:observations}. The results are shown in Sect.~\ref{sec:results} 
and discussed in Sect.~\ref{sec:discussion}. Finally, we give our conclusions 
in Sect.~\ref{sec:conclusions}.

\section{Observations and data reduction \label{sec:observations}}

Observations at 8.35~GHz and with 1.1~GHz receiver bandwith were made 
using the single-horn receiver in  the
secondary focus of the 100-m Effelsberg radio telescope\footnote{The
100-m telescope at Effelsberg is operated by the Max-Planck-Institut f\"ur
Radioastronomie (MPIfR) on behalf of the  Max-Planck-Gesellschaft.}.
The galaxy has been observed by making the maps (called coverages), 
scanned alternatively in R.A. and Dec. A total of 22 coverages was
obtained. Four data channels were recorded.
The first two channels contain total power signals. The correlations of
the left- and right-handed circular polarization signals 
(giving Stokes Q and U) are recorded in the other two channels.
At 4.85 GHz we used the two-horn system in the secondary focus of the
Effelsberg telescope with the receiver of 0.5~GHz bandwith. Four data channels (two Stokes I as well as Q
and U -- see above) were recorded for each horn.  We obtained 10
azimuth-elevation coverages of  NGC\,4569.

The telescope pointing was checked at time intervals of about 2 hours by
making cross-scans of nearby strong point sources. The flux density scale
was calibrated by mapping the highly polarized source 3C286 and computing
its total power flux densities of 4.47~Jy at 8.35~GHz and 7.44~Jy at 4.85~GHz
using the formulae of Baars et al. (\cite{baars}). The same
calibration factors were used for total power and polarized intensity, which
yields a mean degree of polarization of 3C286 of 11\% at 8.35~GHz and
10.5\% at 4.85~GHz, in agreement with published values
(Tabara \& Inoue \cite{tabara}).

The data reduction was performed using the NOD2 data reduction package
(Haslam \cite{haslam}). At 8.35~GHz all the total power coverages were
combined into the final Stokes I map using the spatial frequency weighting
method (Emerson \& Gr\"ave \cite{emgrav}). At 4.85~GHz we combined the
total power information from both horns, using the "software
beam-switching"  technique (Morsi \& Reich \cite{morsi}). This was
followed by a restoration of total  intensities (Emerson et al. \cite{emerson}),
map transformation to R.A./Dec. coordinates and spatial frequency-weighted
combination leading to the final Stokes I map.

Because of azimuthal mounting of the radio telescope the Q and U data at
both frequencies were corrected for the rotation of the telescope reference
frame of polarization with respect to the sky. At 8.35~GHz the  distributions of
Stokes parameters were combined into final Q and U maps using the same
technique as for total power data. At 4.85~GHz  the Stokes U and Q data for
each coverage from both horns were averaged, then rotated to the R.A./Dec.
frame and  combined into final Q and U maps. A digital filtering process,
which removes spatial frequencies  corresponding to noisy structures smaller than
the telescope beamwidth, was  applied to final maps of all the Stokes
parameters at both frequencies. The Q and U maps were finally converted
into maps of polarized intensity, accounted for the "positive noise bias" in the
way described by Wardle \&  Kronberg~(\cite{wardle}). We also computed
the distributions of {\em apparent} (i.e. uncorrected for Faraday rotation --
small in our case) polarization B-vectors, defined as $\mathrm{arctan(U/Q)}+90\degr$.

Instrumental polarization of the Effelsberg radio telescope forms specific
"butterfly patterns" in Q and U maps at the level of about 1\% of an
unpolarized signal (see Klein et al. \cite{klein} for  examples).  They yield an
almost axisymmetric structure in polarized intensity  and in orientation of
polarization vectors, constant in the reference frame of the telescope.
Corrections for the azimuthal mounting makes the vectors of instrumental
polarization rotated by the parallactic angle. In case of observations at a wide
range of parallactic angles (which is our case) the superposition of spurious
polarization patterns with various angles reduces the influence of
instrumental polarization by factor of several. With the
unpolarized intensities discussed in further Sections any instrumental
polarization is in our case well below the noise.

The r.m.s. noise levels in total and polarized intensity at 8.35~GHz are
0.25~mJy/b.a. and  0.045~mJy/b.a., respectively. At 4.85~GHz the corresponding
noise levels are 0.4~mJy/b.a. in total intensity and 0.09~mJy/b.a. in polarized
intensity.

At 1.4~GHz we used the continuum residuals after subtracting the
H{\sc i} line signal from the observations by Vollmer et al.
(\cite{vollm2}). Because of a low sensitivity (bandwidth of only 3 MHz) a
convolution to the beamwidth of $65\arcsec\times 50\arcsec$ was applied to
show the low-surface brightness structure. No polarization information was
recorded at this frequency. The r.m.s. noise level in this map is
0.3~mJy/b.a.

\section{Results \label{sec:results}}

\subsection{The global distribution of radio emission at 4.85~GHz}

\begin{figure}
\resizebox{\hsize}{!}{\includegraphics{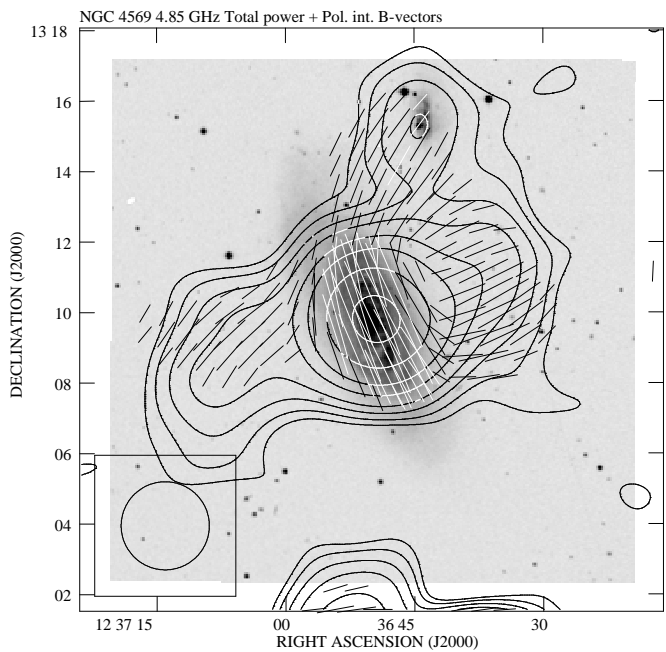}}
\caption{
The total power map of NGC\,4569 at 4.85~GHz with superimposed 
B-vectors of polarized intensity, overlaid upon the blue image from the
DSS. The contour levels are (3, 5, 8, 12, 20, 30, 50, 80) $\times$ 
0.4 (r.m.s. noise level)~mJy/b.a. The polarization vector of 1\arcmin \, 
corresponds to a polarized intensity of 0.5~mJy/b.a. The angular resolution 
is 2\farcm 5.
}
\label{tp6}
\end{figure}

\begin{figure}
\resizebox{\hsize}{!}{\includegraphics{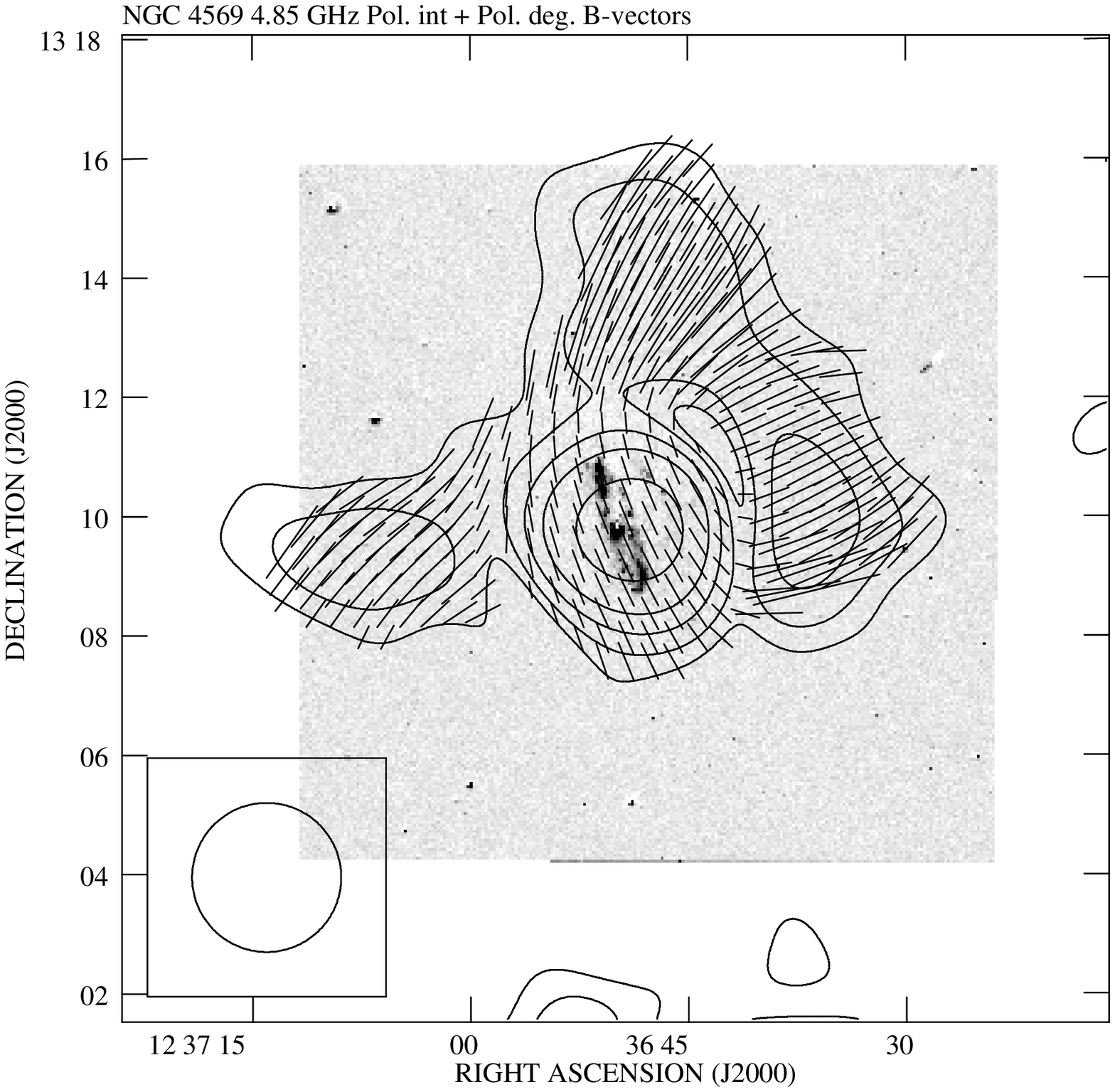}}
\caption{
The contour map of polarized intensity of NGC\,4569 at 4.85~GHz with
superimposed  B-vectors of polarization degree, overlaid upon the 
H$\alpha$ image taken from GOLDMine database (Gavazzi et al. \cite{gavazzi}). 
The contour levels are (3, 5, 8, 12, 20, 30, 50, 80) $\times$ 
0.09 (r.m.s. noise level)~mJy/b.a. The polarization vector of 1\arcmin \,  corresponds to the 
polarization degree of 20\%. The angular resolution is  2\farcm 5. 
}
\label{pi6}
\end{figure}

In contrast to the X-ray emission (Tsch\"oke et al. \cite{tschoke}) the total 
power brightness at 4.85~GHz shows large extensions on both sides of the galactic
disk (Fig.~\ref{tp6}). These large structures were totally unexpected.
In the eastern lobe the radio surface brightness initially decreases shortly 
with distance up to about 6\arcmin ($\sim 30$~kpc) and then quickly drops. 
The western lobe is coincident with the X-ray emission and shows a more 
gradual decrease in surface brightness. Its extent is smaller than the 
eastern lobe (4\arcmin $\sim 20$~kpc). At the position of the dwarf galaxy 
IC~3583 bright total power emission is seen.

The disk of NGC\,4569 and both lobes show significant polarization at 
4.85~GHz (Fig.~\ref{tp6}, \ref{pi6}). In the disk the B-vectors 
(i.e. E-vectors rotated by 90\degr) are parallel to the disk plane, but in 
the extended radio structures the B-vectors are mostly perpendicular to 
the disk. The mean polarization degree in the disk is 5\% increasing to 
15\% in the extraplanar radio-emitting regions. The radio emission from 
IC~3583 is polarized as well, with a degree of polarization of about 12\%. 
The B-vectors are oriented towards NGC\,4569. 

The polarized intensity in the eastern lobe forms a structure elongated 
perpendicularly to the disk, only barely resolved along its minor axis 
(Fig.~\ref{pi6}). In contrast, the polarized emission on the western 
side extends smoothly along the whole disk edge. Even taking into account 
the beam-smeared contribution from IC~3583 leaves the polarization 
considerably more extended along the disk on the western than on the 
eastern disk side. 

\subsection{Details of the radio structure at 8.35~GHz and 1.4~GHz}

\begin{figure}
\resizebox{\hsize}{!}{\includegraphics{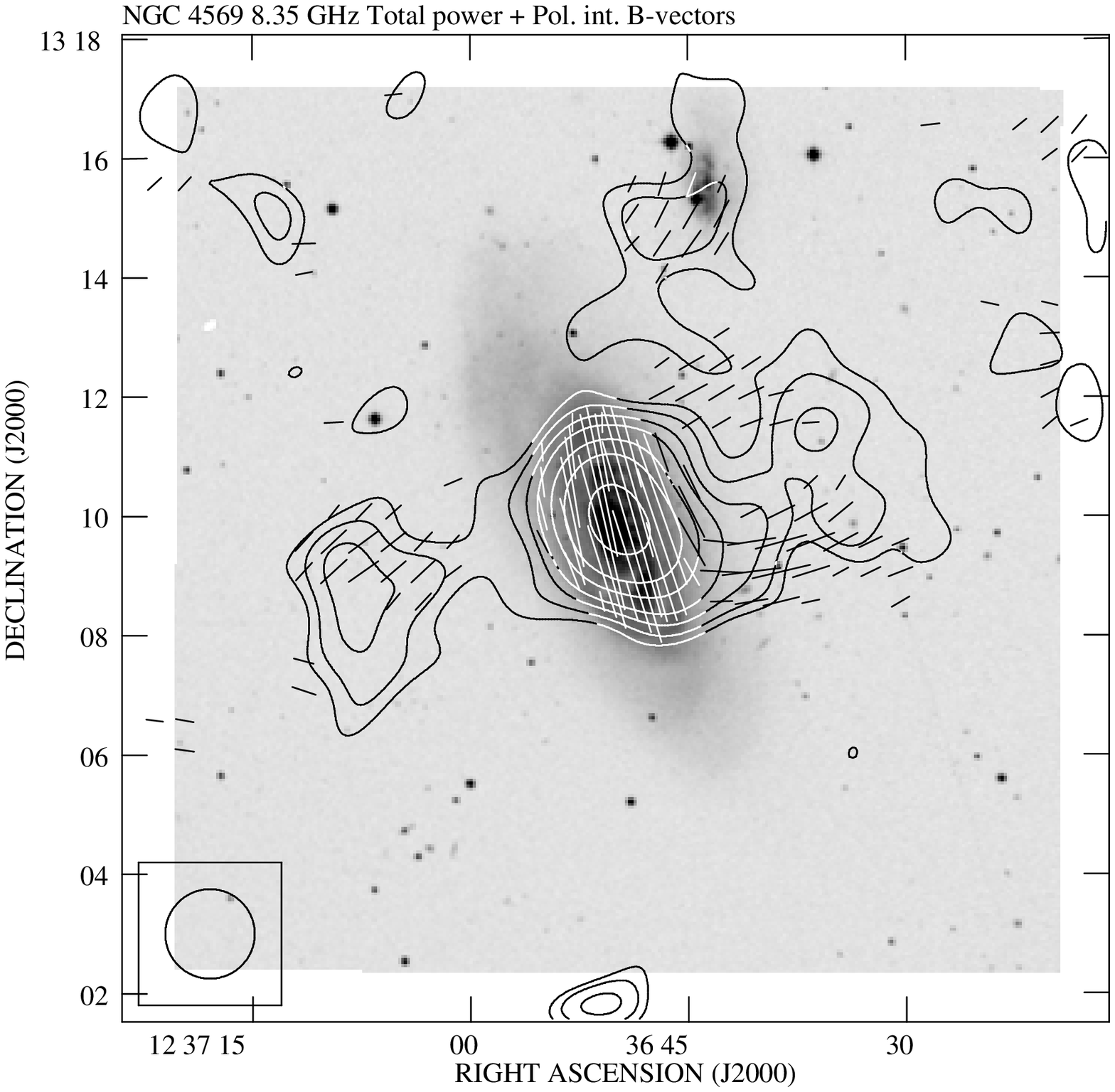}}
\caption{
The total power map of NGC\,4569 at 8.35~GHz with superimposed 
B-vectors of polarized intensity, overlaid upon the blue image from the 
DSS. The contour levels are  (3, 5, 8, 12, 20, 30, 50, 80) 
$\times$0.25 (r.m.s. noise level)~mJy/b.a. The polarization vector of 1\arcmin \, 
corresponds to a polarized intensity of 0.5~mJy/b.a. The angular resolution 
is 1\farcm 5.
}
\label{tp3}
\end{figure}

\begin{figure}
\resizebox{\hsize}{!}{\includegraphics{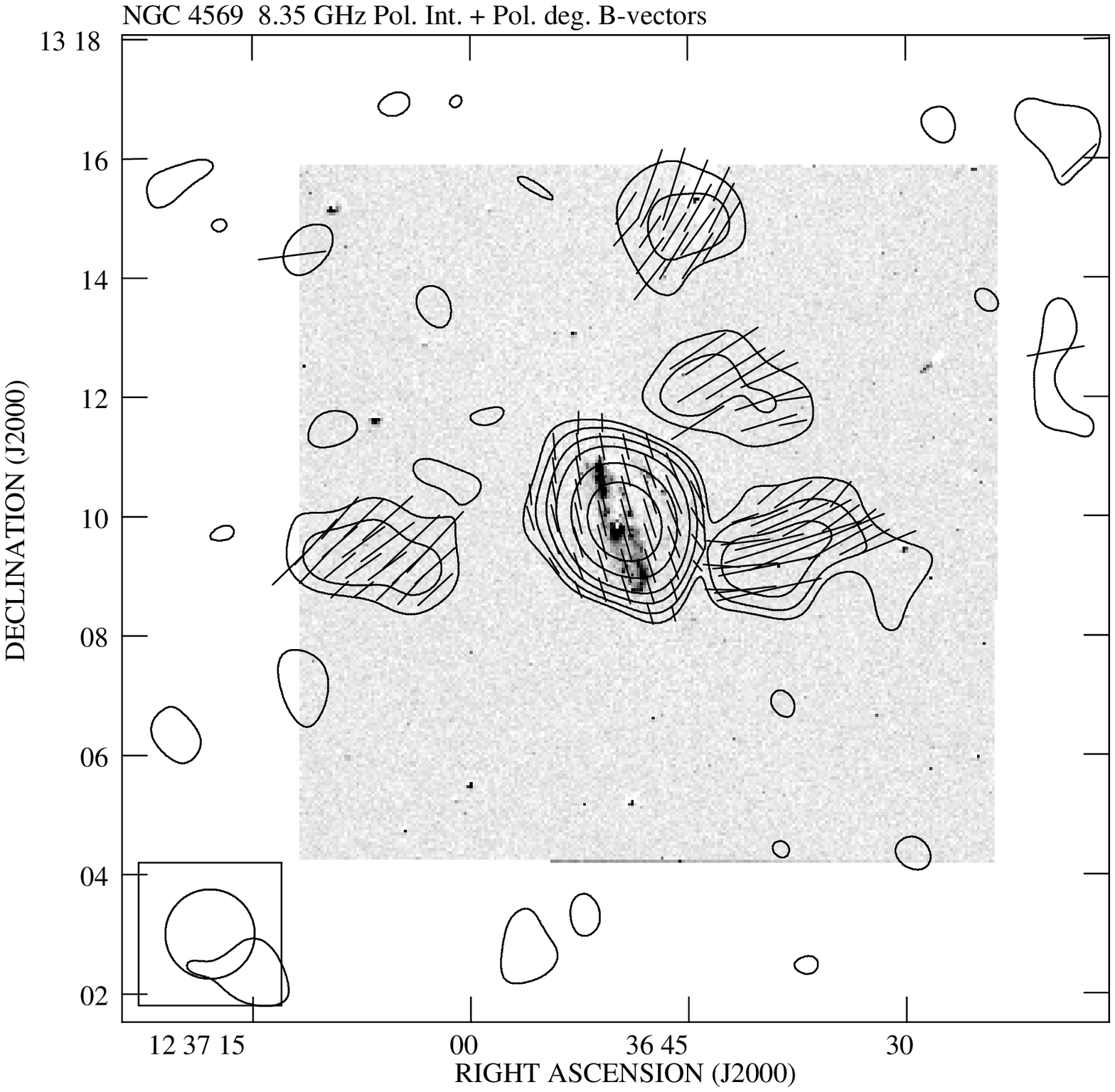}}
\caption{
The contour map of polarized intensity of NGC\,4569 at 8.35~GHz with
superimposed  B-vectors of polarization degree, overlaid upon the 
H$\alpha$ image (from GOLDMine database (Gavazzi et al. \cite{gavazzi})). 
The contour levels are (3, 5, 8, 12, 20, 30, 50, 80) 
$\times$ 0.045 (r.m.s. noise level)~mJy/b.a. The polarization vector of 1\arcmin \,  
corresponds to a polarization degree of 20\%. The angular resolution is 1\farcm 5. 
}
\label{pi3}
\end{figure}

\begin{figure}
\resizebox{\hsize}{!}{\includegraphics{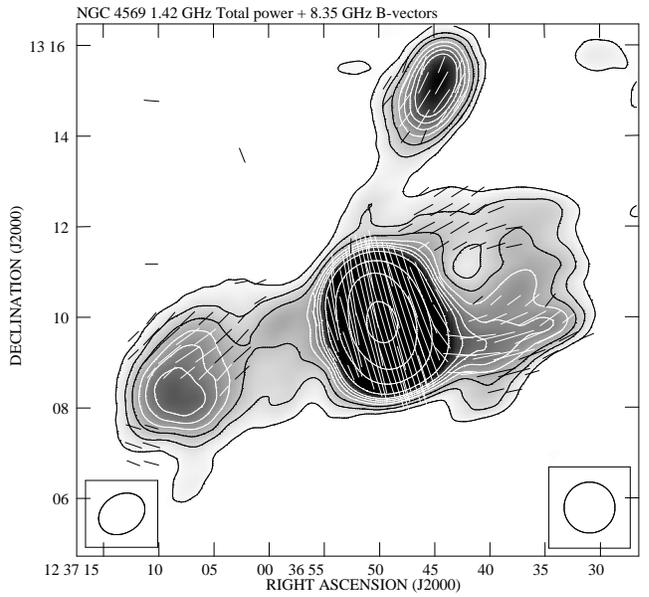}}
\caption{
Combined image showing the distribution of total power emission at
1.49~GHz (contours and greyscale) and orientations of the B-
vectors at 8.35~GHz. The total power map was made from continuum
residuals left after subtraction of the H{\sc i} signal from the
observations of Vollmer et al. (\cite{vollm2}). The B-vectors are from the
observations with the Effelsberg radio telescope. 
The contour levels are 1, 1.5, 2, 2.5, 3, 3.5, 5, 10, 20,
40~mJy/b.a. The r.m.s. noise level is 0.3~mJy/b.a. The beamwidths of the 
total power and polarization maps are shown in the left and right lower 
corner, respectively. The angular resolution is $65\arcsec \times 50\arcsec$.
 }
\label{tp20}
\end{figure}

Our maps at 8.35~GHz with an angular resolution of 1\farcm 5 show the details 
of the total power and polarization structure of NGC\,4569 (Fig.~\ref{tp3}). 
The eastern lobe appears to be a detached region with some hints of a radio 
bridge connecting it to the disk. Only its northern part is polarized by some 
10\% with B-vectors almost perpendicular to the disk. The western lobe has a 
similar shape in total power. However, it is somewhat weaker at 8.35~GHz than 
the eastern region. On this image IC~3583 is separated from NGC\,4569.

Our observations at 8.35~GHz show that the polarized emission on the western 
and NW disk side is concentrated in three separated features (Fig.~\ref{pi3}):  
the most northern polarized peak is associated with IC~3583 and the ridge of 
polarized emission seen at 4.85~GHz west of the disk appears to be composed 
of two regions at the distance of about 3\arcmin \, from the centre of 
NGC\,4569. The fact that they apparently coincide with regions showing little 
total power emission is due to a five times higher noise level in total power 
channels  (due to confusion with background sources and an unpolarized 
atmospheric noise). We can thus detect significant  polarization even where 
the total power signal is weak and very noisy. Both polarization peaks west 
of the disk show B-vectors perpendicular to the disk plane. 

The map at 1.4 GHz made from the line-free channels of the H{\sc i}
observations by Vollmer et al. (\cite{vollm2}) has the highest 
angular resolution of our data, but contains no polarization channels. The 
eastern lobe seems again to be a detached structure, a thin radio bridge 
connecting it to the disk of NGC\,4569 is also confirmed (Fig.~\ref{tp20}). 
The western extension has a more complex structure and seems to be connected 
to the disk. In its southern part a relatively narrow ridge of total power 
emission extends away from the disk. It coincides roughly with the southern 
polarized extension seen west of the disk at 8.35~GHz (Fig.~\ref{pi3}). 
Bright radio emission from IC~3583, which might be connected to that of 
NGC\,4569 via another putative faint radio bridge, is also visible in 
Fig.~\ref{tp6} and \ref{tp3}. The possible interactions between the galaxies
has already been discussed by Tsch\"oke et al. (\cite{tschoke}) mentioning 
the faint southern H$\alpha$ spur, seen in the H$\alpha+$NII image of IC\,3583,
pointing towards NGC\,4569. This is confirmed by a new Fabry Perot 
H$\alpha$ image and velocity field obtained for these galaxies (Chemin et al. 
\cite{chemin}).

\section{Discussion \label{sec:discussion}}

\subsection{Spectral index map}

\begin{figure}
\resizebox{\hsize}{!}{\includegraphics{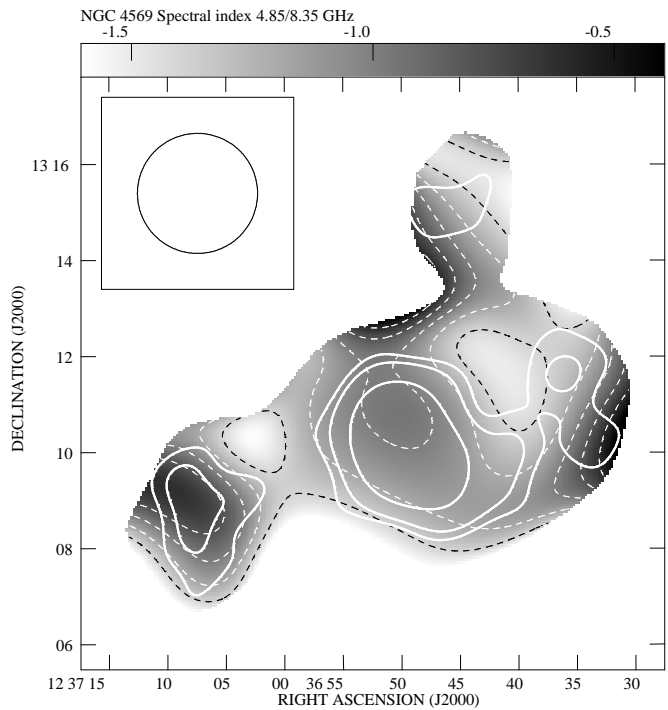}}
\caption{
The distribution of spectral index in NGC\,4569 between 4.85 and 
8.35~GHz from our Effelsberg observations. All the data are convolved to a 
common beamwidth of 2\farcm 5. The dashed contours of spectral index are (-
1.40, -1.20, -1, -0.800, -0.600. Thick solid contours delineate the brightest 
features from the full-resolution map at 8.35~GHz.
}
\label{spix}
\end{figure}

Our results clearly indicate the presence of an unexpected bipolar outflow 
from the inner disk of NGC\,4569. To determine the history of relativistic 
electrons in the extended lobes we computed the distribution of the spectral 
index $\alpha$ ($S_\nu \propto \nu ^{\alpha}$) between 8.35~GHz and 4.85~GHz 
over the whole source structure (Fig.~\ref{spix}). The total power maps at both 
frequencies were convolved to a common beamwidth of 2\farcm 5. The disk has 
a rather steep spectrum with a slope of about $-1.0$. The diffuse western 
lobe has an even steeper spectrum with a slope steeper than $-1.3\, \div\,-1.4$. 
The steepening of the spectral index is due to the energy loss of the 
relativistic electrons via synchrotron emission. Since electrons with higher 
energies lose their energy more rapidly, the spectrum steepens if no 
re-acceleration of the electrons takes place. Thus the observed steepening 
of the spectral index in the western lobe is due to the aging of the 
relativistic electrons. Apparent flattenings at the most extreme western edge 
and at the northern disk boundary are found in regions of a very weak signal 
and may be artifacts of local background variations (a deviation of 1.5 r.m.s
value in the 8.35~GHz map may explain the observed flattening). Acceleration of
the cosmic rays in the southern part of the western lobe, with slightly 
flatter spectral index, cannot be excluded, but high resolution radio data
would be needed to clarify this.

On the other hand, the region of flat spectrum ($\sim-0.6$) in the eastern 
lobe is certainly real. The slope of about $-0.6$ is only slightly steeper 
than the injection spectrum of relativistic electrons in a strong non-magnetic 
shock in a non-relativistic gas with compression ratio of 4 (Beck \& Krause 
\cite{beck05}). Since no tracers of ionized gas are found in the existing 
H$\alpha$ data in this region, an increased thermal fraction is unlikely, 
thus in-situ electron acceleration in a large-scale shock must take place.

At 8.35~GHz only 70\% of the total flux density comes from the disk, the 
remaining 30\% being emitted by the extended structures. Similar fractions are 
found at 1.49~GHz. Thus, a substantial amount of magnetic and cosmic ray energy 
is expelled from the galaxy to the intergalactic space. Such vertical nuclear 
outflows are known to exist in some disk galaxies with strong nuclear 
starbursts or AGN activity, like M82 (Reuter et al. \cite{reuter}), the 
Circinus Galaxy (Elmouttie et al. \cite{circinus}) or NGC~4258 (Krause et al. 
\cite{krause}). All these mentioned objects have extremely powerful central 
sources, dominating their radio emission. In NGC\,4569 the data at 1.49~GHz 
processed with the highest possible resolution (beamwidth of 37\arcsec 
$\times$ 20\arcsec = 3~kpc $\times$ 1.6~kpc) show that the unresolved central 
source comprises no more than 15~mJy, thus only some 10\% of the total flux 
density. High resolution 6~cm and 20~cm data revealed a flux density of 
1.5~mJy at 6~cm and 10~mJy at 20~cm in the inner $4'' = 330$~pc (Neff \& 
Hutchings \cite{neff}, Hummel et al. \cite{hummel}). According to Gabel \& 
Bruhweiler (\cite{gabel}) half of the flux density at 6~cm is due to thermal 
electrons. They suggest that the rest of the flux density is due to 100-200 
supernova remnants.

The eastern lobe has some characteristics of a hot spot at the terminal point 
of the jet expelled from a possible AGN in the centre of NGC\,4569: is this
evidence for a local cosmic ray acceleration and a possible connection (jet) 
to the disk. The available data exclude the existence of a present-day AGN in 
the center of NGC\,4569 (see e.g. Barth \& Shields \cite{barth00} or Gabel \& 
Bruhweiler 2002). Instead, NGC\,4569 harbors a low-ionization nuclear 
emission-line region (LINER). Thus its nucleus shows a very low activity. 
Furthermore, the western lobe does not resemble the classical AGN-expelled 
structure. It is a diffuse feature attached to the disk with vertical magnetic 
fields emerging from almost the whole inner disk. The possibility of an 
earlier AGN activity is discussed in the next section.

\subsection{A galactic wind in a ram pressure wind}

In this section we discuss various possible origins of the observed 
radio-bubbles in NGC\,4569. We start with geometrical aspects of the radio 
emission and an AGN scenario. We then analyse time scales for galactic outflows 
and galactic superwinds. For galactic superwinds we estimate energy needed to 
blow-up radio lobes and confront it with star formation in the disk and with 
two episodes of nuclear starburst. Finally we discuss the influence of cluster 
environments on the development of a ridge of radio continuum emission in the 
southern part of the western lobe.

\subsubsection{The radio geometry}
\label{radgeom} 

A very surprising aspect of the radio lobes is their relative symmetry.
NGC\,4569 is moving rapidly within the hot intracluster medium. According
to Vollmer et al. (\cite{vollm2}) the galaxy has passed the cluster core 
$\sim 300$~Myr ago and the current ram pressure is about 
$p_{\rm ram} \sim 2 \times 10^{-12}$~g\,cm$^{-1}$s$^{-2}$. Moreover, the 
galaxy is moving to the north-east. If the observed radio lobes were due to 
an AGN, one would expect the formation of a head-tail radio galaxy, because 
AGN jets are very vulnerable against forces perpendicular to the outflow 
direction, which is the case for a ram pressure wind. In contrast, in a 
galactic outflow/superwind a pressure driven bubble is created which expands
in all directions and is thus much more stable against forces parallel to the 
galactic disk. We therefore suggest that the radio lobes are not due to an AGN
(even now extinct), but due to a starburst-induced galactic outflow/superwind.

\subsubsection{Galactic outflows}

We cannot yet determine reliably if the material in the flow leaves the 
potential of NGC\,4569 (which is the defining property of a galactic wind) or 
if some or most of the material returns to the disk of NGC\,4569 (which is 
typical for a galactic outflow). The radial velocity of the H$\alpha$ emitting 
gas cone at a radius of 4 kpc is measured to be about 100 km s$^{-1}$ 
(Bomans et al. \cite{bomans05}). The derived time needed to blow-up our radio 
lobe ($25-30$~kpc in size) is then about $\sim 200$~Myr, which is too long 
a life-time for the synchrotron electrons which do not show gradual spectral 
steepening. On the other hand it is possible that the magnetic field and cosmic 
rays are connected to a much hotter phase of the gas forming a superwind-like 
flow at considerably higher velocity. 

\subsubsection{Superwinds}

Typical outflow velocities of galactic superwinds are $\sim 700 -
1000$~km\,s$^{-1}$ (Heckman et al. \cite{heckman}). So $\sim 30$~Myr are
needed to develop the observed spatial extent of the radio lobes. In
order to  balance ram pressure, the inner pressure of the lobes has to
be greater or  equal to the external ram pressure $p_{\rm in} \ge p_{\rm
ram}$. Assuming the minimum energy condition we computed the total
pressure of cosmic rays and magnetic fields in the lobes. A mean value
inside regions delineated by the level of 1.2~mJy/b.a. at 4.85~GHz is
$\approx 1-1.3\times 10^{-12}$dyn\,cm$^{-2}$. It rises to $\ge 2\times
10^{-12}$dyn\,cm$^{-2}$ when brighter parts of lobes inside the level
of  4~mJy/b.a. are considered. Similar values are obtained when pressure
balance conditions are assumed. These values are comparable in range to
the ram pressure (see Sect.~\ref{radgeom}).
The joint magnetic and cosmic-ray energy density is also
of order of 1.5 -- $2\times 10^{12}$~erg~cm$^{-3}$, again similar
to the kinetic energy density of the wind. The outflows have thus a
good chance to overcome the ambient gas pressure while the intergalactic
wind can still deform the extended lobes (Sect.~\ref{ridge}).
 
To estimate the total energy requirements we assume a cone-like
geometry for the outflow with a height of 25~kpc and a diameter of
15~kpc, the total volume occupied by the outflow is thus $V \sim 
8\times 10^{67}$~cm$^{3}$. The minimum energy needed to
overcome the external ram pressure within this region then is $E \sim p_{\rm
in} V \ge p_{\rm ram} V=10^{56}$~erg. With a typical  supernova II total
kinetic energy input $E_{\rm SN}=10^{51}$~erg we find that  the total
number of supernovae driving the outflow is $N_{\rm SN} \ge 10^{5}$. The
mass ejected by the wind-type flow is $M_{\rm wind} = \rho_{\rm wind} V$,
where $\rho_{\rm wind}$ is the mean gas density within the lobes. Assuming
a  mean outflow velocity of $v_{\rm wind}=700$~km\,s$^{-1}$ the mean
density is $\rho_{\rm wind} \sim 2 E/{V v_{\rm wind}^{2}}=5.1 \times
10^{-28}$~g~cm$^{-3}$ and the ejected mass is M$_{\rm wind} \sim 1.6 \times
10^{7}$~M$_{\odot}$. The fraction between the gravitational and the
kinetic energy density is $\varepsilon_{\rm grav}/\varepsilon_{\rm in} 
\sim \big(v_{\rm rot}/v_{\rm wind}\big)^{2}
\sim 0.1$, where we assumed a symmetric dark matter halo and a constant
rotation  velocity of $v_{\rm rot}=250$~km\,s$^{-1}$. Thus, a
sufficiently strong  superwind-type flow can easily overcome the
gravitational potential of the  galaxy and resist ram pressure.

\subsubsection{Disk star formation and the central starbursts}

Is the star formation activity of NGC\,4569 strong enough to explain such
powerful wind-type flows? The disk of NGC\,4569 is weakly forming stars. Due
to the small star formation rate the arm-interarm contrast in the disk is
low and the galaxy is classified as anemic by van den Bergh (\cite{vdB}).
Thus the star forming H$\alpha$ disk (e.g. Koopmann et al. \cite{koopmann})
of NGC\,4569, truncated by the recent ram pressure stripping
event, cannot provide the necessary energy input to drive the observed radio
lobes.

We are thus left with central starbursts. IUE ultraviolet and
optical  ground-based spectra of the nucleus of NGC\,4254 of various
dispersions were  studied by Keel (\cite{keel}). Narrow Balmer
absorption lines and the overall shape of optical spectrum indicates that
the optical light is dominated by a young starburst
particularly rich in A-type supergiants. UV light is more
peculiar and could not be modelled by the UV light from
a dominant population of A supergiants. It was interpreted
as an AGN  or an additional unusually compact and extraordinary luminous
central-star cluster. This last possibility was confirmed in the
detailed study of HST/FOC UV data by Gabel \& Bruhweiler (\cite{gabel}).
Their spectral synthesis and photoionization analyses imply that this
central starburst is very recent ($5-6$~Myr old) and consist of about $5
\times 10^{4}$ O and B stars packed in a region of $\approx 30$~pc in
size. They also pointed out that the population of A supergiants
discerned by  Keel (\cite{keel}) must be older ($>15$~Myr) and
by a factor of 10 more extended than the UV-bright core.
This may mean that its content of OB stars (hence supernova progenitors)
could be in the past considerably greater than $10^5$.

The central present day starburst is clearly too young
($5-6$~Myr) to be responsible  for huge observed radio lobes (timescale
of $\sim 30$~Myr). Also the  number of OB stars generated by the recent
starburst is less than the minimum number of supernovae II ($\sim
10^{5}$) needed to drive the observed radio structures.
In contrast to that the timescale of the second, older
starburst ($>15$~Myr), seen now as an abundant
population of A-type  supergiants matches the timescale of the observed outflow.
This makes this old starburst
(possessing at the past time also enough OB stars)
better candidate to drive the observed
radio lobes.  However, a further detailed modelling of A
supergiant-dominated population  is needed to explain together with the
OB dominated compact starburst the both UV and  optical spectra and to
constrain the starburst geometry, age and energetics, and to model the
radio lobes formation. A detailed stellar population
synthesis  modeling of a past starburst
which would match the present-day
A-star content and produce enough OB stars
at the time of its youth is required to state
whether such an event is sufficient to energize
the observed lobes.

\subsubsection{Radio ridge and ram pressure stripping}
\label{ridge}

The horizontal ridge of radio continuum emission visible at 1.49~GHz in the 
southern part of the western lobe of NGC\,4569 (Fig.~\ref{tp20}) coincides 
with a spot of polarized emission (Fig.~\ref{pi3}) parallel to this structure.
The western side of the galactic disk is the side closest to the Virgo Cluster 
centre. In the ram pressure scenario of Vollmer et al. (\cite{vollm2}) 
NGC\,4569 is moving within the Virgo intracluster medium to the north-east. 
Consequently, the ISM of NGC\,4569 is pushed by ram pressure to the south-west 
(last frame of Fig.~6 of Vollmer et al. \cite{vollm2}). Thus, ram pressure is 
not directly responsible for the  horizontal ridge of radio continuum emission.

We suspect the horizontal ridge of radio continuum emission to be outflowing 
gas of the galactic superwind colliding with the part of the stripped ISM 
of NGC\,4569 which is still close to the galaxy. If this gas is atomic its
surface density must be smaller than $10^{20}$~cm$^{-2}$, because it was not
detected in the VLA data of Vollmer et al. (\cite{vollm2}). On the other hand,
if this putative south-western H{\sc i} plume is more extended than 20-30~kpc,
its H{\sc i} mass cannot exceed several $10^{7}$~M$_{\odot}$, because it would 
have been detected with the Effelsberg 100-m telescope. The collision of 
outflowing and stripped gas leads to compression of the outflowing gas and 
the magnetic field contained in it, giving rise to the observed horizontal 
ridge of enhanced total power and polarized radio continuum emission. In 
addition, we expect that a shock is formed when the wind is hitting the 
relatively dense stripped ISM of NGC\,4569. The observed flattening
of the spectral index in the south-western part of the western radio lobe
might then be due to in-situ particle acceleration in this large-scale shock.

The flat  spectral index of the eastern lobe and the implied in-situ
acceleration of relativistic electrons there might be due to a large-scale 
shock driven by the direct impact of ram pressure on this lobe, whereas the 
western lobe is protected from a direct impact of ram pressure by the 
stripped ISM of NGC\,4569. This shielding might also be partly responsible for 
the different aspects of the two radio lobes. High resolution radio continuum 
observations are needed to confirm and resolve the distribution of the 
spectral index in the radio lobes. If our scenario is correct we would 
expect a flatter spectral index in the direction of the ram pressure wind, 
i.e. to the north-east in the eastern lobe.

Of course, our scenario has caveats: (i) why is the stripped and partly
re-accreting gas seen in Fig.~6 of Vollmer et al. (\cite{vollm2}) located 
more to the west of the galaxy center than to the south-west? (ii) Can such 
gas, if it exists, efficiently shield/protect the galactic wind-like flows? 
Of what type is the stellar population, its age and extent responsible for
radio lobes? More sensitive multi-frequency radio polarization and optical 
observations of NGC\,4569 with high spatial resolution would greatly help to 
establish a detailed scenario of these peculiar gas outflows.

\section{Conclusions \label{sec:conclusions}}

Using the Effelsberg radio telescope working at 4.85~GHz and 8.35~GHz we 
discovered large scale outflows in the strongly stripped Virgo spiral galaxy 
NGC\,4569. We found that:
\begin{itemize}
\item[-] NGC\,4569 possesses large radio lobes extending up to 24~kpc 
from the disk, which is unusual for normal spirals and even more unusual
for cluster spirals.
\item[-] The lobes emit as much as 30\% of the total flux density at 8.35~GHz.
\item[-] The eastern lobe shows signatures of in-situ shock-driven cosmic ray 
acceleration. 
\item[-] The western lobe looks like a diffuse halo with vertical magnetic 
fields spread over the whole disk.
\item[-] Using the continuum channels at 1.4~GHz from the H{\sc i} 
observations we suggest that its southern part shows effects of gas and 
magnetic field compression by the ambient intracluster medium.
\end{itemize}

The radio structure of NGC\,4569 differs in many respects from that of
normal  spiral galaxies: the radio lobes with scales of
tens of kiloparsecs are very rare phenomena in field spiral galaxies
and have not yet been observed in a cluster spiral
galaxy. The  extraplanar western ridge of unpolarized and polarized
radio continuum  emission is also a peculiar feature, which is most
probably due to the  cluster environment.

The available data exclude present-day and past AGN activity in the core
of  NGC\,4569. We exclude that the observed radio lobes are due to an
extinct AGN,  because NGC\,4569 undergoes relatively strong ram pressure
effects (Vollmer et al. \cite{vollm2}) which would lead to a head-tail galaxy.
Instead, we propose  that the observed radio lobes are due to a
galactic superwind-like flows,  induced by a starburst lasting several 
$10$~Myr. Such  an event, which requires $10^5$ supernovae
explosions and $10^{56}$~erg  total input energy, is consistent with
a stellar population dominated by a large number of A
supergiants in the central region of the galaxy (Keel 1996). This is supported
by estimates of the combined magnetic and cosmic-ray pressure inside the
lobes from our radio data.

Our observations give the first evidence that galactic superwind-like flows can 
occur even in a spiral galaxy located near the cluster center (projected 
distance $=0.5$~Mpc.) They can expand to distances of several 10~kpc and might 
finally escape the galaxy's gravitational potential and enrich the 
intracluster medium with gas, dust and magnetic fields. In addition to 
magnetic fields expelled from galaxies during their interactions (Chy\.zy 
\& Beck \cite{chyzy04}) objects like NGC\,4569 may constitute an important 
source of intergalactic/intracluster magnetic fields, thereby relaxing the seed field 
problem in the dynamo theories (e.q. Widrow \cite{widrow02}).

Despite this first evidence, we are far from fully understanding the nature of
the giant outflows in NGC\,4569. High resolution, high sensitivity, 
multi-frequency radio continuum observations including polarization will 
greatly help to test our tentative scenario. 
The polarized radio emission detected from IC\,3583 and the existence of 
a radio bridge connecting IC\,3583 and NGC\,4569 suggest a possible 
interaction between the two objects which will have to be addressed, too.

\begin{acknowledgements}
The authors wish to express their thanks to the colleagues from the 
Max-Planck-Institut f\"ur Radioastronomie (MPIfR) in Bonn for their valuable 
discussions during this work. KCh, MS,  and MU are indebted to Professor 
Richard Wielebinski from the MPIfR for the invitations to stay at 
this institute, where substantial parts of this work were done. This work 
was supported by a grant from the Polish Research Committee (KBN), 
0249/P03/2001/21. DJB acknowledges the SFB\,591 `Universal Beheavior of
non-equilibrium plasmas'. We have made use of the LEDA and GOLDMine databases.
\end{acknowledgements}

{}

\end{document}